\PassOptionsToPackage{table,xcdraw}{xcolor}
\documentclass[11pt,a4paper]{article}
\pdfoutput=1

\usepackage{geometry}
\usepackage[numbers,sort&compress]{natbib}
\usepackage{amsmath}
\usepackage{amssymb}
\usepackage[dvips]{graphicx}
\usepackage{pstricks}
\usepackage{bm}
\usepackage{pbox}
\usepackage{placeins}
\usepackage{graphicx}
\usepackage{caption}
\usepackage{subcaption}
\usepackage[T1]{fontenc}
\usepackage{footnote}
\usepackage{pdfpages}
\usepackage{hhline}
\usepackage{multirow}
\usepackage{enumitem}
\usepackage{slashed}
\usepackage[nottoc,notlot,notlof]{tocbibind}
\usepackage[title,titletoc]{appendix}
\usepackage{dsfont}
\usepackage{pifont}
\allowdisplaybreaks
\usepackage{cases}
\usepackage{tikz}
\usepackage[symbol]{footmisc}
\usepackage{physics}

\renewcommand{\thetable}{\Roman{table}}

\def\equationautorefname~#1\null{Eq.\,(#1)\null}

\definecolor{MyDarkBlue}{rgb}{0.1, 0.1, 0.8}
\definecolor{MyLightBlue}{rgb}{0.22,0.51,0.9}
\definecolor{MyGreen}{rgb}{0.0, 0.5, 0.0}
\definecolor{BrickRed}{rgb}{0.8, 0.25, 0.33}
\usepackage[colorlinks=true,linkcolor=blue,citecolor=MyDarkBlue,
urlcolor=MyLightBlue,bookmarksnumbered=true,bookmarksopen]{hyperref}
\hypersetup{colorlinks, citecolor=blue,linkcolor=blue, urlcolor=MyLightBlue}

\begin{document}
\vspace*{-0.2in}
\begin{flushright}

\end{flushright}
\vspace{0.5cm}
\begin{center}
{\Large\bf
Fermion mass, Axion dark matter, and Leptogenesis\\ in SO(10) GUT
}
\end{center}

\vspace{0.5cm}
\renewcommand{\thefootnote}{\fnsymbol{footnote}}
\begin{center}
{\large
{}~\textbf{Ajay Kaladharan$^1$}\footnote[1]{ E-mail: \textcolor{MyLightBlue}{kaladharan.ajay@okstate.edu}}, 
{}~\textbf{Shaikh Saad$^2$}\footnote[2]{E-mail:  
\textcolor{MyLightBlue}{shaikh.saad@unibas.ch}
}
}
\vspace{0.5cm}

{\em $^1$Department of Physics, Oklahoma State University,\\ Stillwater, OK 74078, USA}
\\
{\em $^2$Department of Physics, University of Basel,\\ Klingelbergstrasse\ 82, CH-4056 Basel, Switzerland}
\end{center}

\renewcommand{\thefootnote}{\arabic{footnote}}
\setcounter{footnote}{0}
\thispagestyle{empty}
\begin{abstract}
SO(10) grand unified theory with minimum parameters in the Yukawa sector employs the Peccei-Quinn symmetry that solves the strong CP problem. Such an economical Yukawa sector is highly appealing and has been extensively studied in the literature. However, when the running of the renormalization group equations of the Yukawa couplings are considered, this scenario shows somewhat tension with the observed fermion masses and mixing. In this work, we propose an extension of the minimal framework that utilizes lower dimensional representations and alleviates this tension by introducing only a few new parameters. The proposed model consists of a fermion in the fundamental and a scalar in the spinorial representations. While the latter is needed to implement the Peccei-Quinn symmetry successfully, the presence of both is essential in obtaining an excellent fit to the fermion mass spectrum. In our model, axions serve the role of dark matter, and the out-of-equilibrium decays of the right-handed neutrinos successfully generate the matter-antimatter symmetry of the Universe.  
\end{abstract}
\newpage
{\hypersetup{linkcolor=black}
\tableofcontents}
\setcounter{footnote}{0}

\section{Introduction}
Grand Unified Theories (GUTs) aim to unify the strong, weak, and electromagnetic forces into a single force at a high energy scale. Several important GUT models~\cite{Pati:1973rp,Pati:1974yy, Georgi:1974sy, Georgi:1974yf, Georgi:1974my, Fritzsch:1974nn}, including those proposed by Pati and Salam~\cite{Pati:1973rp,Pati:1974yy} as well as by Georgi and Glashow~\cite{Georgi:1974sy}, have been extensively studied in the literature.

One particularly intriguing class of GUTs is based on the SO(10) symmetry~\cite{Georgi:1974my, Fritzsch:1974nn}. What makes these models fascinating is their ability to accommodate all Standard Model (SM) fermions within a single irreducible 16-dimensional representation. Notably, this 16-dimensional spinorial representation includes the SM singlet right-handed neutrino. Consequently, these models can also account for the tiny masses of the Standard Model (SM) neutrinos through the type-I seesaw mechanism~\cite{Minkowski:1977sc,Yanagida:1979as,Glashow:1979nm,Gell-Mann:1979vob,Mohapatra:1979ia}. Moreover, since the GUT symmetry can break down to the SM gauge group via multiple intermediate stages, gauge coupling unification can be obtained without requiring light states. 

Within the renormalizable SO(10) framework, the Higgs representations that can contribute to the fermion masses and mixings can be determined by the following fermion bilinear:
\begin{equation}
16\times 16 = 10_s+ 120_a + 126_s,
\end{equation}
where subscripts $s$ and $a$ represent symmetric and antisymmetric components (in the family space). With the above tensor product, the Yukawa Lagrangian takes the general form,
\begin{equation}
\mathcal{L}_{yuk}= 16_F (Y_{10}10_H+Y_{120}120_H+Y_{126}\overline{126}_H) 16_F.
\end{equation}
Among the $3\times 3$ Yukawa coupling matrices, $Y_{10}, Y_{126}$ are symmetric, whereas $Y_{120}$ is antisymmetric in the family space.  The Yukawa sector of SO(10) GUTs is remarkably predictive and has undergone thorough analysis in numerous studies~\cite{Babu:1992ia, Bajc:2001fe,Bajc:2002iw,Fukuyama:2002ch,Goh:2003sy,
Goh:2003hf,Bertolini:2004eq, Bertolini:2005qb, Babu:2005ia,Bertolini:2006pe, Bajc:2008dc,
Joshipura:2011nn,Altarelli:2013aqa,Dueck:2013gca, Fukuyama:2015kra, Babu:2016cri, Babu:2016bmy, Saad:2017wgd, Babu:2018tfi, Babu:2018qca, Mohapatra:2018biy, Ohlsson:2018qpt, Ohlsson:2019sja,Babu:2020tnf,Mummidi:2021anm,Saad:2022mzu,Haba:2023dvo}.

The most minimal Yukawa sector, consistent with only SO(10) gauge symmetry, is proposed in Ref.~\cite{Babu:2016bmy}. On the other hand, additional symmetries can be imposed to further reduce the number of parameters. One such well-motivated version is extending the theory by a global Peccei-Quinn (PQ) symmetry~\cite{Peccei:1977hh,Peccei:1977ur} that solves the strong CP problem~\cite{Peccei:1977hh,Peccei:1977ur,Weinberg:1977ma,Wilczek:1977pj,Kim:1979if,Shifman:1979if,Zhitnitsky:1980tq,Dine:1981rt}. In this scenario, with a complex $10_H$ and a $\overline{126}_H$ Higgs representations, the Yukawa sector consists of the minimum number of parameters as the PQ symmetry forbids one of the two Yukawa terms with $10_H$. The validity of such a minimal Yukawa sector has been shown in several works by performing numerical fits, e.g., Refs.~\cite{Joshipura:2011nn,Babu:2018tfi}. Due to the many orders of difference between the electroweak (EW) and GUT scales, one must carefully consider the renormalization group equations (RGEs) running of the relevant Yukawa couplings during the fitting procedure. Such analysis has been performed, e.g., in Refs.~\cite{Dueck:2013gca,Ohlsson:2018qpt,Ohlsson:2019sja,Saad:2022mzu}.

As shown in Refs.~\cite{Dueck:2013gca,Ohlsson:2018qpt,Ohlsson:2019sja}, once RGE running (which includes the threshold effects of the right-handed neutrinos) is incorporated properly, the minimal Yukawa sector with $Y_{10}$ and $Y_{126}$ is unable to fit all observables in the charged fermion and the neutrino sectors within $2\sigma$ ranges. The best fit presented in 
Ref.~\cite{Dueck:2013gca} (Ref.~\cite{Ohlsson:2019sja}) with a total $\chi^2=22.97$ (14.8) has the maximum pull for the top-quark mass ($\theta_{23}$ mixing angle in the neutrino sector), which is $\sim -3.4$ ($\sim -2.4$). See also Ref.~\cite{Ohlsson:2018qpt}, which adopted a slightly different approach to the RGE running and found much higher deviations (total $\chi^2=85.9$). A crucial thing to note is that in the analysis of Refs.~\cite{Dueck:2013gca,Ohlsson:2019sja}, 
for observable with higher precision than 5$\%$, a rather large uncertainty of 5$\%$ is set. Instead, taking much smaller values of the uncertainties closer to their experimental values is likely to show tension of higher degrees. Furthermore, adding baryon asymmetry parameter in the fit is expected to result in even larger deviations of these observables from their measured values.

In this work, we propose a simple extension of the minimal model that alleviates the tensions in the fermion mass fit, as mentioned above. In particular, we allow only lower dimensional representations and introduce a fermion in the fundamental representation, $10_F$, and a scalar in the spinorial representation, $16_H$. This proposed model introduces only a limited number of additional parameters in the Yukawa sector with which an excellent fit to fermion masses and mixings can be obtained. Once the $16_H$, the presence of which is necessary to implement the PQ symmetry successfully, gets a vacuum expectation value (VEV), the fermion $10_F$ mixes with the usual fermions, $16_i$, which modifies the mass matrices of the light SM-like fermions and help to obtain better fits. Since PQ symmetry is expected to be broken at an intermediate state that sets the scale for right-handed neutrinos, light axions appear that can accommodate the entirety of the dark matter. Moreover, through the out-of-equilibrium decays of the heavy right-handed neutrinos, the matter-antimatter asymmetry of the Universe can be incorporated.

This paper is organized in the following way. In Sec.~\ref{sec:model}, we introduce the proposed model and work out the details of the Yukawa sector. Numerical fit is performed in Sec.~\ref{sec:numerics} and details of the PQ mechanism is described in Sec.~\ref{sec:PQ}. In Sec.~\ref{sec:leptogenesis}, we discuss how matter-antimatter asymmetry of the Universe is computed. Finally, we conclude in Sec.~\ref{sec:conclusions}.

\section{Model}\label{sec:model}
\subsection{Yukawa sector}
With a complex $10_H$ and a $\overline{126}_H$, one has the following set of Yukawa interactions:
\begin{equation}
\label{yukawa}
\mathcal{L}_{yuk}\supset 16_F (Y_{10}10_H+Y_{126}\overline{126}_H) 16_F.
\end{equation}
Since $10_H$ is taken to be complex, a second Yukawa coupling associated with it, $16_F16_F10_H^\ast$ is also allowed in general~\cite{Babu:1992ia}. This additional term is typically forbidden by imposing a PQ symmetry, $U(1)_\mathrm{PQ}$. The introduction of the PQ symmetry is motivated since it is needed in order to solve the strong CP problem. Under $U(1)_\mathrm{PQ}$,  the Higgses $10_H$ and $\overline{126}_H$ carry negative two units and  the fermion $16_i$ carries positive one unit of charge.

With the above Yukawa coupling \autoref{yukawa}, the fermion mass spectrum, in the $f^T M_f f^c$ basis, is determined by the following matrices:
\begin{align}
&M_u=vY_u=v_u^{10}Y_{10}+v_u^{126}Y_{126}, \\
&M_d=vY_d=v_d^{10}Y_{10}+v_d^{126}Y_{126},\\
&M_e=vY_e=v_d^{10}Y_{10}-3v_d^{126}Y_{126},\\
&M^D_\nu=vY^D_\nu=v_u^{10}Y_{10}-3v_u^{126}Y_{126},\\
&M_R=v_RY_{126},
\end{align}
with $v=174.104$ GeV. Here, we denote the up-type and down-type EW VEVs of the $10_H$ ($126_H$) as $v^{10}_u$ ($v^{126}_u$) and $v^{10}_d$ ($v^{126}_d$), respectively. Moreover, the VEV of the SM singlet field within $\overline{126}_H$ is represented by $v_R$. The above set of matrices can be rewritten as,
\begin{align}
&Y_d= H + F,\\
&Y_u= r (H + s F),\label{up-mass}\\
&Y_e= H -3 F,\\
&Y^D_\nu= r (H - 3 s F),\\
&M_R=c_R F,
\end{align}
\noindent where we have defined the following quantities:
\begin{align}
&Y_{10}=\frac{v}{v^{10}_{d}} H, \;\;\;
 Y_{126}=\frac{v}{v^{126}_{d}} F,
 \;\;\; r= \frac{v^{10}_u}{v^{10}_d}, \;\;\; s= \frac{v^{126}_u}{v^{126}_d} \frac{v^{10}_d}{v^{10}_u}, \;\;\; c_R= v_R \frac{v}{v^{126}_d}.
\end{align}
Moreover, the light neutrino masses are determined by the type-I seesaw,
\begin{align}
M_\nu= -v^2 Y^D_\nu M_R^{-1}\left(Y^D_\nu\right)^T \;\label{nu-mass}.    
\end{align}

As mentioned in the introduction, we propose to add a fermion, $10_F$, and a scalar, $16_H$, to alleviate the tensions in the fermion masses within this minimal setup. Furthermore, the breaking of the GUT symmetry to an intermediate symmetry is  performed by $54_H$ dimensional representation. 
Alternatively, one could add a multiplet with higher dimensional representation, $120_H$ (instead of a $10_F$ and a  $16_H$), which has a direct Yukawa coupling with the fermions. However, in such a scenario, another multiplet needs to be added to consistently break the PQ symmetry at the high scale to guarantee invisible axions (and not EW scale axions). This second multiplet is expected to play no role in fermion mass spectrum. Therefore, our proposed model is more attractive since it not only utilizes lower dimensional representations but also both the multiplets participate non-trivially in correcting the fermion masses and mixings.

We assign the following charges to these representations under the PQ symmetry ($i=1,2,3$):
\begin{align}
\mathrm{Fermions:}\;\;\;&
16_F^i\to  e^{+i\alpha}   16_F^i,
\;\;\;
10_F\to   10_F.
\\
\mathrm{Scalars:}\;\;\;&
10_H\to  e^{-2i\alpha}   10_H,
\;\;\;
\overline{126}_H\to  e^{-2i\alpha}   \overline{126}_H,
\;\;\;
54_H\to   54_H,
\;\;\;
16_H\to  e^{-i\alpha}   16_H. \label{PQ-charge}
\end{align}

First, note that mass of the quark-like states ($D, D^c$) and lepton-like states ($E, E^c$ and $N, N^c$) residing in $10_F$ have independent masses, 
\begin{align}
\mathcal{L}_Y&\supset 
10_F10_F\left(m_F+y\; 54_H \right)
\\
&= 
\underbrace{(2m_F+2\sqrt{2}y v_{54})}_{\equiv m_F^\prime} D^cD
+\underbrace{(2m_F-3\sqrt{2}y v_{54})}_{\equiv m_F^{\prime\prime}}   \left(EE^c+NN^c\right)\;.
\end{align}
Furthermore, with the above charge assignments, one obtains mixings between $16_F$ and $10_F$\begin{align}
\mathcal{L}_Y&\supset 
z_i16_i10_F16_H= \underbrace{- \sqrt{2} z_i  v_{16}}_{\equiv \mu_i} \left(d_i^cD+e_iE^c-\nu_i N^c\right),
\end{align}
where, $v_{16}\equiv \langle 16_H\rangle$. Since $z_i$ are Yukawa couplings, demanding perturbative couplings, one expects $\mu_i \lesssim v_{16}$.  Consequently, we derive the following the $4\times 4$ Dirac mass matrices,
\begin{align}
\mathcal{L}_Y&\supset \begin{pmatrix}
d_i&D    
\end{pmatrix} M_D \begin{pmatrix}
d_i^c\\D^c    
\end{pmatrix} 
+\begin{pmatrix}
e_i&E    
\end{pmatrix} M_E \begin{pmatrix}
e_i^c\\E^c    
\end{pmatrix}
+\begin{pmatrix}\nu_i&N    
\end{pmatrix} M_N^D \begin{pmatrix}
\nu_i^c\\N^c    
\end{pmatrix},
\end{align}
with
\begin{align}
&M_D=\begin{pmatrix}
M_d&0_{3\times 1}\\
\mu_{1\times 3}&m_F^\prime
\end{pmatrix},\;\;\;
M_E=\begin{pmatrix}
M_e&  \mu^T_{3\times 1}\\
0_{1\times 3}&m^{\prime\prime}_F
\end{pmatrix},\;\;\;
M^D_N=\begin{pmatrix}
M^D_\nu&  -\mu^T_{3\times 1}\\
0_{1\times 3}&m^{\prime\prime}_F
\end{pmatrix}\;. \label{full}
\end{align}

Finally, integrating out the heavy fermions leads to $3\times 3$ matrices of the light fermions (in the $f^T M_f f^c$ basis), namely up-type quarks, down-type quarks, charged leptons, and Dirac neutrinos, respectively 
\begin{align}
&Y_u^\mathrm{light}=Y_u, \label{eq:1}
\\
&Y_d^\mathrm{light}=Y_d\left(1+r^\dagger_D r_D \right)^{-1/2}, \label{eq:1d}
\\
&Y_e^\mathrm{light}=\left(1+r^T_E r^*_E \right)^{-1/2}Y_e, \label{eq:1e}
\\
&Y_{\nu_D}^\mathrm{light}=\left(1+r^T_E r^*_E \right)^{-1/2}Y^D_\nu, \label{eq:2}
\end{align}
where we have defined,
\begin{align}
&r_D\equiv  \left(r_1\;r_2\;r_3\right),\;\;\;
r_E\equiv  r_0\left(r_1\;r_2\;r_3\right),
\\
&\left(r_1\;r_2\;r_3\right)\equiv \frac{1}{m_F^\prime}\left(\mu_1\;\mu_2\;\mu_3\right), \;\;\;   
r_0\equiv \frac{m^\prime_F}{m^{\prime\prime}_F}.
\end{align}
The light neutrino mass matrix is then obtained from, 
\begin{align}
M_\nu= -v^2 Y_{\nu_D}^\mathrm{light} M_R^{-1}\left(Y_{\nu_D}^\mathrm{light}\right)^T \;\label{eq:3}.    
\end{align}
Therefore, in the proposed model, the fermion mass matrices are given by \autoref{eq:1}-\autoref{eq:2} and \autoref{eq:3}.

Here we clarify that after integrating out the heavy states{\color{black},} our obtained \autoref{eq:1d}-\autoref{eq:2} are valid as long as $M_d\ll \mu, m^\prime_F$   and $M_e, M^D_\nu \ll \mu, m^{\prime\prime}_F$~\cite{Babu:1995uu}. Our derivation is quite general, and does not require that $r_{D,E}$ have to be smaller than unity. Therefore, the $3\times 3$ effective Yukawa/mass matrices derived above are in excellent agreement with the quantities computed from the full  $4\times 4$ matrices. In Appendix~\ref{A}, we explicitly demonstrate this by computing eigenvalues from both $3\times 3$ and $4\times 4$ matrices. However, if the vectorlike fermions are somewhat light, i.e., if they have masses close to the TeV (or below), one must diagonalize the entire $4\times 4$ to accurately determine the eigenvalues and eigenvectors --a case we do not consider in this work.

\subsection{Symmetry breaking}
The complete symmetry of our model is $SO(10)\times U(1)_\mathrm{PQ}$ and 
the charge assignments of the Higgs fields are presented in \autoref{PQ-charge}.  Since $54_H$ is uncharged under $U(1)_\mathrm{PQ}$, its VEV does not break the PQ symmetry. The $54_H$ field spontaneously breaks the GUT symmetry to the Pati-Salam symmetry with a preserved D-parity~\cite{Chang:1983fu}. In principle, the $\overline{126}_H$ field can break this Pati-Salam symmetry to the SM gauge group. This breaking, however, would leave a linear combination of $U(1)_X\subset SO(10)$ and
$U(1)_\mathrm{PQ}$ unbroken, see, e.g., Ref.~\cite{Babu:2018qca}. To consistently break the PQ symmetry and realize only $U(1)_Y$ at low energies requires another symmetry breaking field with a non-trivial PQ charge. The lowest dimensional representation to achieve this is a spinorial representation. If the VEVs of $16_H$ and $\overline{126}_H$ are of similar order, then the symmetry breaking chain in our model is given by 
\begin{align}
SO(10)\times U(1)_\mathrm{PQ} 
&\xrightarrow[54_H]{M_\mathrm{GUT}} 
SU(4)_{C}\times SU(2)_{L} \times SU(2)_{R}\times D \times U(1)_\mathrm{PQ}   \label{breaking-1}
\\
&\xrightarrow[16_H+\overline{126}_H]{M_\mathrm{int}} 
SU(3)_{C}\times SU(2)_{L} \times U(1)_{Y} 
\\
&\xrightarrow[10_H+\overline{126}_H]{M_{EW}} 
SU(3)_{C} \times U(1)_\mathrm{em}\;. \label{breaking-3}
\end{align}

The first (and the second) symmetry breaking produces superheavy monopoles, which must be diluted not to overclose the Universe~\cite{Kibble:1976sj,Linde:1981mu}.  Moreover, spontaneous breaking of the PQ symmetry (along with the non-perturbative  QCD effects) leads to multiple distinct degenerate vacua resulting in $N_\mathrm{DW}$ number of domain walls, leading to the so-called axion domain wall problem~\cite{Sikivie:1982qv}. Therefore, we assume inflation~\cite{Guth:1980zm,Albrecht:1982wi,Linde:1981mu,Linde:1983gd} (that can be achieved via a gauge singlet field) to take place after the scale $M_\mathrm{int}$ (but before the leptogenesis scale, i.e., $M_\mathrm{int}> M_2$, which can be easily arranged), which gets rid of all unwanted topological defects.

On the other hand, if the VEVs of $54_H$ and $16_H$ are taken to be at the GUT scale, then one gets,
\begin{align}
SO(10)\times U(1)_\mathrm{PQ} 
&\xrightarrow[54_H+16_H]{M_\mathrm{GUT}} 
SU(3)_{C}\times SU(2)_{L} \times U(1)_{Y} \times U(1)_{\mathrm{PQ}^\prime}   \label{breaking-4}
\\
&\xrightarrow[\overline{126}_H]{M_\mathrm{int}} 
SU(3)_{C}\times SU(2)_{L} \times U(1)_{Y}
\\
&\xrightarrow[10_H+\overline{126}_H]{M_{EW}} 
SU(3)_{C} \times U(1)_\mathrm{em}\;, \label{breaking-5}
\end{align}
where, in the first stage, an Abelian global symmetry,   which we denote by $U(1)_{\mathrm{PQ}^\prime}$, still remains unbroken (see discussion above). In both these scenarios, $\langle \overline{126}\rangle\sim M_\mathrm{int} \ll M_\mathrm{GUT}$ is required to give correct masses to neutrinos. To get rid of all unwanted topological defects, as before, we require inflation to take place after the second stage of the symmetry breaking.  Although it may be possible to achieve inflation utilizing one of the Higgses from the symmetry-breaking sector (this, however, requires special conditions to be satisfied by the relevant potential), one may alternatively employ a scalar, singlet under the GUT gauge group, as the inflaton (see, for example, Ref.~\cite{Pallis:2016mvm}). However, the details of the inflation dynamics are irrelevant to our study.

In this work, our focus is the newly proposed Yukawa sector; hence, we do not provide the details of gauge coupling unification. Note, however, that with a minimal number of relevant fields, a GUT scale of order $M_\mathrm{GUT}\sim 2\times 10^{15}$ GeV can be obtained with the symmetry breaking chain \autoref{breaking-1}-\autoref{breaking-3} (see, e.g., Ref.~\cite{Babu:2016bmy}).  In this scenario, the GUT symmetry is first broken to the Pati-Salam group with the discrete D parity intact. In the final step, the Pati-Salam gauge group is broken down into the SM gauge group. Using the low scale measured values of the gauge couplings, one finds the $SU(2)_L$ and $SU(2)_R$ gauge couplings to unify ($g_L=g_R$) at $5\times 10^{10}$ GeV scale. Furthermore, from the minimal survival hypothesis, assuming the presence of bi-doublets form $10_H$ and $\overline{126}_H$, as well as $(10,3,1)+(10,1,3)$ from $\overline{126}_H$ to reside at the intermediate scale, a unification scale of order $\sim 10^{15}$ GeV can be obtained~\cite{Babu:2016bmy}. By taking into account the threshold correction from the scalars, one can easily obtain a larger GUT scale (see, e.g., Ref.~\cite{Saad:2022mzu} for details)  to be consistent with the current proton decay bounds that require $M_\mathrm{GUT}\gtrsim 6\times 10^{15}$ GeV (see, e.g., Ref.~\cite{Dev:2022jbf}). Since the exact value of the proton decay lifetime cannot be computed without delving into the details of the calculation, we comment that, as in Ref.~\cite{Babu:2016bmy}, we expect the $p\to e^+\pi^0$ and $p\to \overline{\nu}\pi^+$ to be the two most dominant decay modes (a characteristic feature of non-supersymmetric $SO(10)$ GUTs). The detailed study of gauge coupling unification and proton decay computation in our model is left for future work. As usual, the doublet-triplet Higgs splitting is obtained by fine-tuning the relevant parameter.

\section{Numerical analysis}\label{sec:numerics}
The fermion mass matrices, as represented by \autoref{eq:1}-\autoref{eq:2} and \autoref{eq:3}, are characterized by a constrained set of parameters. Specifically, there are 16 magnitudes and 10 phases to reproduce 19 observables. These observables are: 6 quark masses, 3 quark mixing angles, 1 CKM phase, 3 charged lepton masses, 2 neutrino mass squared differences, 3 mixing angles in the neutrino sector, and the baryon asymmetry parameter $\eta_B$.  We exclude the Dirac CP phase in the neutrino sector, as it remains unmeasured to date. 

\FloatBarrier
\begin{table}[th!]
\centering
\footnotesize
\resizebox{0.65\textwidth}{!}{
\begin{tabular}{|c||c|c|c|}
\hline
\textbf{Observables} & \multicolumn{3}{c|}{Values at $M_Z$ scale}  \\ 

\cline{2-4}
($\Delta m^2_{ij}$ in $eV^2$) &Input&Fit &pull$^2$
\\
\hline \hline

\rowcolor{teal!20}$y_u/10^{-6}$&6.65$\pm$2.25&6.55&$1.95\times 10^{-3}$\\  
\rowcolor{teal!20}$y_c/10^{-3}$&3.60$\pm$0.11&3.59&$2.79\times 10^{-6}$\\ 
\rowcolor{teal!20}$y_t$&0.986$\pm$0.0086&0.986&$3.11\times 10^{-3}$\\ 
\hline\hline

\rowcolor{red!13}$y_d/10^{-5}$&1.645$\pm$0.165&1.646&$6.23\times 10^{-5}$\\ 
\rowcolor{red!13}$y_s/10^{-4}$&3.125$\pm$0.165&3.126&$3.87\times 10^{-5}$\\ 
\rowcolor{red!13}$y_b/10^{-2}$&1.639$\pm$0.015&1.639&$3.34\times 10^{-3}$\\ 
\hline\hline

\rowcolor{yellow!13}$y_e/10^{-6}$&2.7947$\pm$0.02794&2.7944&$1.32\times 10^{-4}$\\ 
\rowcolor{yellow!13}$y_\mu/10^{-4}$&5.8998$\pm$0.05899&5.8962&$3.79\times 10^{-3}$\\ 
\rowcolor{yellow!13}$y_\tau/10^{-2}$&1.0029$\pm$0.01002&1.0028&$7.80\times 10^{-5}$\\ 
\hline\hline

\rowcolor{blue!13}$\theta_{12}^\textrm{CKM}/10^{-2}$&$22.735\pm$0.072&22.732&$1.65\times 10^{-3}$\\ 
\rowcolor{blue!13}$\theta_{23}^\textrm{CKM}/10^{-2}$&4.208$\pm$0.064&4.210&$1.25\times 10^{-3}$\\ 
\rowcolor{blue!13}$\theta_{13}^\textrm{CKM}/10^{-3}$&3.64$\pm$0.13&3.64&$1.62\times 10^{-4}$\\ 
\rowcolor{blue!13}$\delta^c_\textrm{CKM}$&1.208$\pm$0.054&1.207&$2.92\times 10^{-4}$\\ 
\hline\hline

\rowcolor{orange!10}$\Delta m^2_{21}/10^{-5} $&7.425$\pm$0.205&7.417&$1.31\times 10^{-3}$\\ 
\rowcolor{orange!10}$\Delta m^2_{31}/10^{-3}$ &2.515$\pm$0.028&2.515&$7.41\times 10^{-5}$\\
\hline\hline

\rowcolor{green!13}$\sin^2 \theta_{12}$ &0.3045$\pm$0.0125&0.3041&$1.14\times 10^{-3}$\\ 
\rowcolor{green!13}$\sin^2 \theta_{23}$   &0.5705$\pm$0.0205 $^\ddagger$&0.4494&$0.59$\\ 
\rowcolor{green!13}$\sin^2 \theta_{13}$ &0.02223$\pm$0.00065&0.02223&$3.31\times 10^{-5}$\\  

\hline\hline
\rowcolor{red!40}$\eta_B/10^{-10}$&6.12$\pm$0.004&6.12&$1.81\times 10^{-4}$\\

\hline \hline
\rowcolor{white!10}$\chi^2$&-&-&0.6 \\
\hline

\end{tabular}
}
\caption{The fitted values of the observables at the low scale for the benchmark fit with $\chi^2=0.6$ (we remind the readers that 19 observables are fitted against 16 magnitudes and 10 phases). ${}^\ddagger$Note that experimental measurements of $\theta_{23}$ have two local minimum~\cite{NUFIT}; although only the best fit from the global fit~\cite{NUFIT} is shown, we have allowed the entire viable ranges in the fitting procedure.  As can be seen from this table, the only significant contribution to the total $\chi^2$ is from $\theta_{23}$. 
}\label{result}
\end{table}

We perform a $\chi^2$-function minimization to this system, where the free parameters are randomly chosen at the GUT scale (which we fix to be $M_{\mathrm{GUT}}=2\times 10^{16}$ GeV). At the GUT scale, the above set of Yukawa/mass matrices is matched with the complete SM + Type-I seesaw RGEs, which are then evolved (using \texttt{REAP}~\cite{Antusch:2005gp}) to the low scale (i.e., to the $M_Z=91.8176$ GeV scale) by successively integrating out the right-handed neutrinos at their respective mass thresholds. For simplicity and following the procedures of Refs.~\cite{Dueck:2013gca,Ohlsson:2019sja}, any corrections to the RGEs from the intermediate scale to the GUT scale due to the presence of the additional states other than right-handed neutrinos are not considered. The inclusion of such modifications, however, is beyond the scope of this work. Finally, we fit the observables at $M_Z$; input values of the observables at this scale are summarized in Table~\ref{result} (see Refs.~\cite{Antusch:2013jca} and~\cite{NUFIT,Esteban:2020cvm}). Since the charged lepton masses and the baryon asymmetry parameter are determined experimentally with great precision, we assume $1\%$ uncertainties for these quantities during the fitting procedure. The $\chi^2$-function is defined as 
\begin{align}
\chi^2= \sum_\mathrm{all\; observables} \left( \frac{ \mathrm{theoretical\; prediction} - \mathrm{experimental\; central\; value}}{\mathrm{experimental}\; 1\sigma \;\mathrm{error}} \right)^2 = \sum \mathrm{pull}^2.    
\end{align}

The parameters at the GUT scale obtained from the fitting procedure described above for a benchmark fit are presented in Appendix~\ref{A}. Moreover, the fit values of the observables are recapitulated in Table~\ref{result}.

For the fit presented in Table~\ref{result}, we find the Dirac CP phase in the neutrino sector to be $\delta_\mathrm{CP}=344.2^\circ$, and the masses of the light neutrinos, as well as the heavy-right handed neutrinos are
\begin{align}
&\left(m_1,m_2,m_3\right)=(0.285, 0.907, 5.02)\times 10^{-11}\; \mathrm{GeV}, 
\\
&\left(M_1,M_2,M_3\right)=
\left( 0.0564, 2.13, 2.37\right)\times 10^{11}\; \mathrm{GeV}.  \label{eq:RHN}
\end{align}

In determining the baryon asymmetry parameter, $\eta_B$, we numerically solve the relevant density matrix equations. 
The details of the computation of leptogenesis are relegated to Sec.~\ref{sec:leptogenesis}. For earlier works on leptogenesis in SO(10) setup, see, e.g. Refs.~\cite{Buchmuller:1996pa,Nezri:2000pb,Buccella:2001tq,Branco:2002kt,Akhmedov:2003dg,DiBari:2008mp,DiBari:2010ux,Buccella:2012kc,DiBari:2013qja,DiBari:2014eya,Fong:2014gea,DiBari:2015oca,DiBari:2017uka,Saad:2017pqj,Chianese:2018rnq,DiBari:2020plh,Patel:2022xxu}.

\begin{figure}[t!]
\centering
\includegraphics[width=0.49\textwidth]{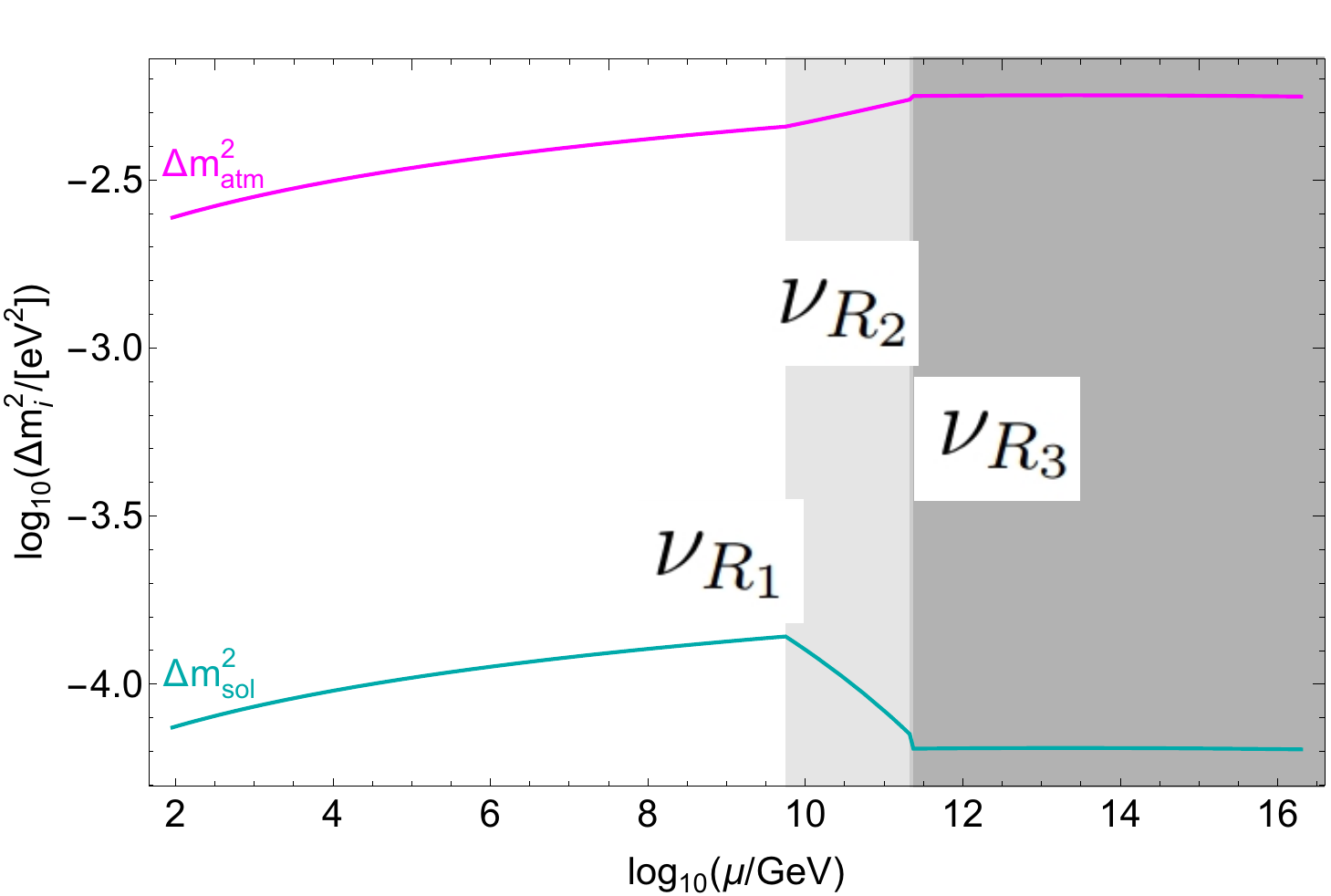}
\includegraphics[width=0.48\textwidth]{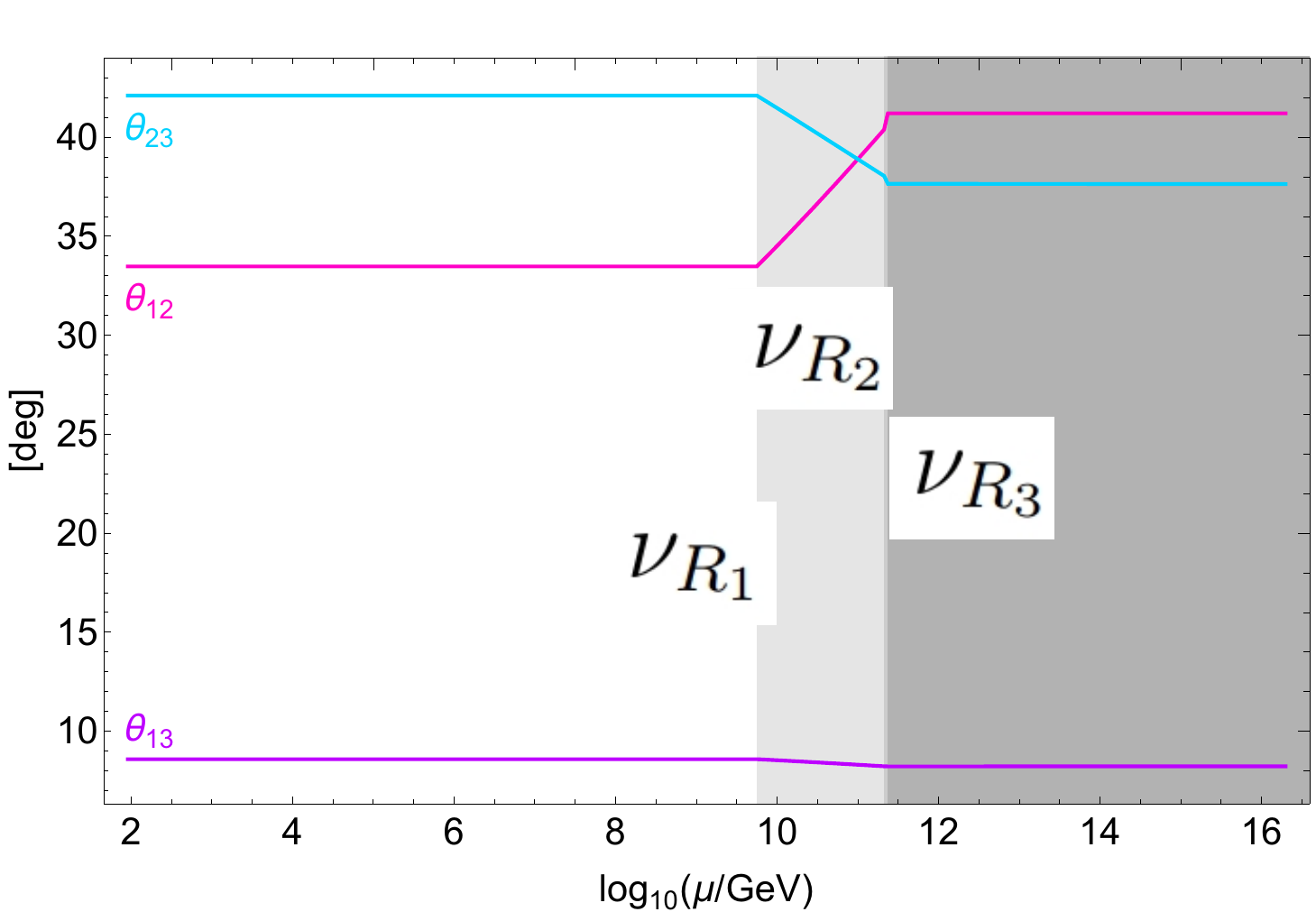}
\caption{  Plots depicting the importance of the RGE running on the  neutrino mass-squared as well as mixing parameters. Scales, where the heavy right-handed neutrinos $\nu_{R_i}$ decouple from the theory, are presented via different shades of gray. See text for details. } \label{fig:RGE}
\end{figure}

The importance of the fitting procedure following the top-down approach that includes the threshold effects due to integrating out the heavy-neutrinos is demonstrated in Fig.~\ref{fig:RGE}. For the benchmark fit presented here, in Fig.~\ref{fig:RGE}, we show the RGE evolution of the neutrino observables from  the GUT scale to the low scale, namely the two mass-squared differences and the three mixing angles in the left and the right panels, respectively. These plots clearly illustrate the difference between the low scale and the high scale values of the observables. For example, for the quantity $\Delta m_\mathrm{atm}^2$, a much larger value compared to the low energy measured value is expected at the GUT scale. Moreover, the two large angles in the leptonic sector, namely $\theta_{12}$ and  $\theta_{23}$  can change significantly due to the decoupling effects of the right-handed neutrinos. These important effects can not be captured in the fitting procedure that follows the bottom-up approach (therefore, cannot include threshold effects from the right-handed neutrinos), which would lead to an inaccurate determination of the model parameters.

\section{Axion dark matter}\label{sec:PQ}
It is crucial to correctly identify the axion, for which we follow Ref.~\cite{Ernst:2018bib}. For this purpose, it is convenient to parameterize the scalars in the following way:
\begin{align}
\phi_k=\frac{1}{\sqrt{2}} \left(\rho_k+v_k\right)e^{ \frac{i A_k}{v_k} },    
\end{align}
where $v_k$ is VEV of the field $\phi_k$. The spontaneous symmetry breaking of the global symmetry leaves a Goldstone, namely the axion in our case, which we denote by $A$. Then one can write, 
\begin{align}
A_k=\underbrace{ \left(\frac{q_k v_k}{f_\mathrm{PQ}}\right) }_{\equiv c_k} A+ \mathrm{orthogonal\;exciations}; \;\;\;\;\; f_\mathrm{PQ}= \left( \sum_k q^2_k v^2_k \right)^{1/2},   
\end{align}
where $q_k$ is the PQ charge of $\phi_k$. Consequently, the axion field is identified as,
\begin{align}
A= \sum_k c_k A_k\;.  \label{axion}  
\end{align}
First, note that the fields that acquire VEVs and carry PQ charges in our setup are given by,
\begin{align}
&10_H\supset (1,2,2)\supset \underbrace{H_u}_{\phi_3}(1,2,1/2)+ \underbrace{H_d}_{\phi_4}(1,2,-1/2) \;,   
\\
&\overline{126}_H\supset (15,2,2)+(10,1,3)\supset \underbrace{\Sigma_u}_{\phi_1}(1,2,1/2)+ \underbrace{\Sigma_d}_{\phi_2}(1,2,-1/2)  + \underbrace{\Delta_R}_{\phi_5}(1,1,0) \;,
\\
&16_H\supset (4,2,1)+(\overline 4,1,2)\supset \underbrace{\xi_d}_{\phi_6}(1,2,-1/2)+ \underbrace{\xi_s}_{\phi_7}(1,1,0) \;.
\end{align}
Since the VEV of $\overline{126}_H\supset (\overline{10},3,1)\supset (1,3,1)$ is super tiny, its contribution can be safely neglected.  

In the following, we identify the axion field by determining the $c_k$ coefficients, for which, first, we apply the orthogonality conditions. This implies that the axion must be orthogonal to the Goldstone bosons of the broken gauge symmetries. Even though the SO(10) group has rank five and has five Cartan generators, the fields that obtain VEVs are color singlet and do not carry electric charge. Hence, only two combinations of the Cartan generators are relevant, which can be taken to be the $U(1)_R$ and $U(1)_{B-L}$. Then, two orthogonality conditions can be found by utilizing,
\begin{align}
\sum_k c_k q^X_k v_k=0,    
\end{align}
where $X=R$ or $X=B-L$, and $q^X_k$ represents the gauge charge of the field $\phi_k$. Correspondingly we obtain, 
\begin{align}
&c_1 v_1 -  c_2 v_2 +  c_3 v_3 -  c_4 v_4 +2\;  c_5 v_5 -  c_7 v_7=0,
\\
&-2\;c_5 v_5 -  c_6 v_6 +  c_7 v_7  =0.
\end{align}

At the perturbative level, the axion  remains massless, which provides additional conditions on $c_k$. Note that the following non-trivial terms are allowed by both gauge and PQ symmetries:
\begin{align}
V\supset 10_H10_H \overline{126}_H^*  \overline{126}_H^* + 16_H 16_H 10^*_H  + 16_H 16_H   \overline{126}_H^* +\mathrm{h.c.} \label{potential}
\end{align}
The first of these terms can be written as,
\begin{align}
V&\supset10_H10_H \overline{126}_H^*  \overline{126}_H^* \supset 
H_u H_u \Sigma_u^*  \Sigma_u^* + H_d H_d \Sigma_d^*  \Sigma_d^*
 \\ 
&\supset -\frac{1}{2}v^2_1v^2_3 \left( \frac{A_3}{v_3}-\frac{A_1}{v_1} \right)^2   -\frac{1}{2}v^2_2v^2_4 \left( \frac{A_4}{v_4}-\frac{A_2}{v_2} \right)^2 , 
\end{align}
which provides the following conditions:
\begin{align}
&\frac{c_3}{v_3}-\frac{c_1}{v_1}=0,\;\;\; 
\frac{c_4}{v_4}-\frac{c_2}{v_2}=0.
\end{align}
Finally, the last two terms in \autoref{potential} leads to,
\begin{align}
V&\supset16_H 16_H 10^*_H  + 16_H 16_H   \overline{126}_H^* \supset
\xi_dH^*_d\xi_s +  \xi_d\Sigma^*_d\xi_s + \xi_s^2 \Delta_R^*
\\& \supset
-\frac{1}{4\sqrt{2}} v_6v_7v_4 \left(   \frac{A_6}{v_6}+\frac{A_7}{v_7}+\frac{A_4}{v_4} \right)^2 
-\frac{1}{4\sqrt{2}} v_6v_7v_2 \left(   \frac{A_6}{v_6}+\frac{A_7}{v_7}+\frac{A_2}{v_2} \right)^2 
-\frac{1}{4\sqrt{2}} v_7^2v_5 \left(  2 \frac{A_7}{v_7}+\frac{A_5}{v_5} \right)^2, 
\end{align}
yielding additional constraints on $c_k$, 
\begin{align}
& \frac{c_6}{v_6}+\frac{c_7}{v_7}+\frac{c_4}{v_4} =0,
\;\;\;
\frac{c_6}{v_6}+\frac{c_7}{v_7}+\frac{c_2}{v_2} =0,
\;\;\;
2 \frac{c_7}{v_7}+\frac{c_5}{v_5}  =0.
\end{align}
The above set of equations for $c_k$ and the requirement for a canonical normalization of the axion provide solutions to these coefficients as follows: 
\begin{align}
&c_1= \frac{x v_1 \left(-6 v_2^2-6 v_4^2+4
   v_5^2+v_7^2\right)}{\left(v_1^2+v_3^2\right)
   v_7}, \;\;\;
c_3= \frac{x v_3 \left(-6 v_2^2-6 v_4^2+4
   v_5^2+v_7^2\right)}{\left(v_1^2+v_3^2\right)
   v_7},
\\ 
&c_2= -\frac{6 x v_2}{v_7}, \;\;\;
c_4= -\frac{6 x v_4}{v_7},  \;\;\;
c_5= -\frac{2 x v_5}{v_7},  \;\;\;
c_6= \frac{5 x v_6}{v_7}, \;\;\;
c_7= \frac{-v_7 \sqrt{v^2_1+v^2_3}}{b_0^{1/2}}\equiv x, \label{eq:b0}
\end{align}
where the expression for the quantity $b_0$ is given in Appendix~\ref{B}. With these coefficients, the axion field is identified from \autoref{axion}, and the physical charges of the scalars are also determined using $\frac{q_k}{f_\mathrm{PQ}}=\frac{c_k}{v_k}$. Moreover, the domain wall number is found to be $N_\mathrm{DW}=6$. As aforementioned, since $N_\mathrm{DW}>1$, inflation must occur after the PQ symmetry breaking to eliminate the domain walls.

Since the PQ symmetry (or the effective PQ symmetry, PQ$^\prime$) is broken at the intermediate symmetry scale, the axion decay constant is roughly given by $f_A\sim M_\mathrm{int}$.  Therefore, the axion mass is determined by~\cite{Sikivie:2006ni},
\begin{align}
    m_A\sim 6\mu\mathrm{eV} \left( \frac{10^{12}\mathrm{GeV}}{M_\mathrm{int}} \right).
\end{align}
Since within our scenario, to get the correct neutrino mass scale, $v_R\sim 10^{12-13}$ GeV is expected,  axion mass scale is predicted to be $m_A\sim \mathcal{O}(1-100)\mu\mathrm{eV}$, which is compatible with both astrophysical and the laboratory experimental bounds~\cite{DiLuzio:2020wdo}, and can be a cold dark matter candidate~\cite{Preskill:1982cy,Abbott:1982af,Dine:1982ah}. The relic abundance of the axion field today can be obtained from~\cite{Sikivie:2006ni}
\begin{align}
    \Omega_A h^2\approx 0.7 \left( \frac{M_\mathrm{int}}{10^{12}\mathrm{GeV}} \right)^{1.16} \left( \frac{\Theta_i}{\pi} \right)^2,
\end{align}
where $\Theta_i$ is the initial misalignment angle of the axion field, which can take values in the range $\Theta_i\subset [-\pi,\pi]$.

We make the following crude estimation  to show that natural values of the initial misalignment angle can incorporate the full dark matter abundance.  Assuming $Y_3$ and $F_3$ to denote the largest entries of $Y_{126}$ and $F$ matrices, respectively, one can write, $M_3\sim v_R Y_3 \sim c_R F_3$, which implies $F_3\sim 1.247\times 10^{-3}$ for our fit.  Moreover, using the definition of the matrix $F$, we further write $Y_3\sim F_3 v/v^{126}_d$. Finally, assuming $v^{126}_d\in (1, 174)$ GeV, we find the viable range for $v_R\sim 10^{12-14}$ GeV. With these values of $v_R\sim M_\mathrm{int}$, correct dark matter relic abundance can be obtained for  $|\Theta_i|\in (0.09, 1.3)$. Here, we have used the dark matter relic abundance  $\Omega_{\textrm{DM}} h^2=0.120\pm 0.001$ as measured by Planck collaboration~\cite{Planck:2018vyg}.

\section{Leptogenesis}\label{sec:leptogenesis}
The lepton asymmetry is generated in thermal leptogenesis by CP violating out-of-equilibrium decays of the right-handed neutrinos. The CP asymmetry occurs via the interference between tree and one-loop diagrams involving the decay of heavy neutrinos into leptons and Higgs. For the right-handed neutrinos with $M_{N_i}\geq 10^{13}\, \mathrm{GeV}$, the leptons $\ket{L_i}$ and anti-lepton $\ket{\bar{L}_i}$ quantum states produced via the decay of $N_{i}$ can be written as pure states between their production at decay and absorption at inverse decay~\cite{Blanchet:2011xq}. On the other hand, for mass regime $10^{12}\,\mathrm{GeV} \geq M_{N_i} \geq 10^{9}\,\mathrm{GeV}$, the coherent evolution of $\ket{L_i}$ and $\ket{\bar{L}_i}$  states break down due to collision with right-handed tauons, before inverse decay can occur~\cite{Blanchet:2011xq}. The lepton $\ket{L_i}$ and anti-lepton $\ket{\bar{L}_i}$ states coupling with $N_i$ can be written in lepton flavour eigenstates $(\alpha=e,\mu, \tau)$ as~\cite{Blanchet:2011xq}, 
\begin{equation}
    \ket{L_i}=\sum_{\alpha}\mathcal{C}_{i\alpha}\ket{L_\alpha}, \quad \mathcal{C}_{i\alpha}\equiv \bra{L_\alpha}\ket{L_i} \quad \quad \mathrm{and}  \quad \ket{\bar{L}_i}=\sum_{\alpha}\mathcal{\bar{C}}_{i\bar{\alpha}}\ket{\bar{L}_{\alpha}}, \quad \mathcal{\bar{C}}_{i\bar{\alpha}}\equiv \bra{\bar{L}_\alpha}\ket{\bar{L}_i}. 
\end{equation}
The CP conjugate of $\ket{\bar{L}_i}$ can be written as,
\begin{equation}
    CP\ket{\bar{L}_i}=\sum_{\alpha}\mathcal{\bar{C}}_{i\alpha}\ket{L_i}, \quad \mathrm{with} \quad  \mathcal{\bar{C}}_{i\alpha}=\mathcal{\bar{C}}^*_{i\bar{\alpha}}. 
\end{equation}
In general $\mathcal{C}_{i\alpha} \neq \mathcal{\bar{C}}_{i\alpha}$ due to one-loop CP violating correction~\cite{Blanchet:2011xq}, but at tree-level they are identical given by,
\begin{equation}
    \mathcal{C}^0_{i\alpha} = \mathcal{\bar{C}}^0_{i\alpha}=\frac {Y_{i\alpha}}{\sqrt{\left (Y^\dagger Y  \right )_{ii}}},
\end{equation}
the matrix $Y$ is defined in Appendix~\ref{B}.

The classical Boltzmann equations cannot capture the asymmetries in the intermediate regime where the lepton quantum states interact with the thermal bath between decay and inverse decay via charged lepton interactions and cannot be represented either as a pure state or as an incoherent mixture. The charge lepton interactions and Yukawa interactions compete to dictate the characters of the lepton quantum states. The density matrix equations are necessary to calculate the asymmetry in this regime~\cite{Blanchet:2011xq}, which are given by~\cite{Blanchet:2011xq, Dev:2017trv}, 
\begin{align}
    \frac {dN_{N_j}}{dz}=&-D_j\left ( N_{N_j}-N_{N_j}^{eq} \right ) \nonumber\\
    \frac {dN^{B-L}_{\alpha \beta}}{dz}=&\sum_j\left [ \varepsilon _{\alpha \beta}^{(j)}D_j\left ( N_{N_j}-N_{N_j}^{eq} \right )-\frac 12W_j\left \{ P^{(j)0},N^{B-L} \right \}_{\alpha \beta} \right ]\nonumber\\
    &-\frac {\mathrm{Im}(\Lambda_\tau )}{Hz}\left [ \begin{pmatrix}
1 & 0 & 0\\ 
0 & 0 & 0\\ 
0 & 0 & 0
\end{pmatrix},\left [ \begin{pmatrix}
1 & 0 & 0\\ 
0 & 0 & 0\\ 
0 & 0 & 0
\end{pmatrix},N^{B-L} \right ] \right ]_{\alpha \beta} \nonumber\\
    &-\frac {\mathrm{Im}(\Lambda_\mu )}{Hz}\left [ \begin{pmatrix}
0 & 0 & 0\\ 
0 & 1 & 0\\ 
0 & 0 & 0
\end{pmatrix},\left [ \begin{pmatrix}
0 & 0 & 0\\ 
0 & 1 & 0\\ 
0 & 0 & 0
\end{pmatrix},N^{B-L} \right ] \right ]_{\alpha \beta},
     \label{eq:dme}
\end{align}
where $z=M_{N_1}/T$, and $N_{N_j}$ ($N^{B-L}$) is the particle number of $N_j$ neutrino ($B-L$ asymmetry) evaluated in the co-moving volume containing one heavy neutrino in ultra-relativistic thermal equilibrium.
The $N_{N_i}^{\mathrm{eq}}$ is the equilibrium number density defined as,
\begin{equation}
N_{N_i}^{\mathrm{eq}}=\frac 12 x_i z^2 \mathcal{K}_2(z_i),
\end{equation}
in order that $N_{N_i}^{\mathrm{eq}}(z_i \simeq 0)=1$. Here $x_i$ and $z_i$ are given by,
\begin{equation}
    x_i=\frac {M_j^2}{M_1^2}, \quad \quad \quad z_i=\sqrt{x_i} z,
\end{equation}
and $\mathcal{K}_i(z)$ is the modified Bessel function of the second kind. The decay term $D_i$ and washout term $W_i$ are given by,
\begin{align}
    D_i&\equiv D_i(z)\equiv \frac {\Gamma_i+\bar{\Gamma}_i}{Hz}=K_i x_i z\frac {\mathcal{K}_1(z_i)}{\mathcal{K}_2(z_i)},\\
W_i&\equiv W_i(z)\equiv \frac 12 \frac {\Gamma_i^{ID}+\bar{\Gamma}^{ID}_i}{Hz}=\frac 14 K_i\sqrt{x_i}z_i^3\mathcal{K}_1(z_i),
\end{align}
where $\Gamma_i$ ( $\bar{\Gamma}_i$) is the decay rate of right-handed neutrino $N_i$ into leptons (anti-leptons), and $\Gamma^{ID}_i$ ( $\bar{\Gamma}^{ID}_i$) is the inverse decay rate of leptons (anti-leptons). The decay parameter $K_i$ is given by,
\begin{equation}
    K_i\equiv\frac {(\Gamma_i+\bar{\Gamma}_i)_{T=0}}{H(M_i)}=\frac {M_i(Y^\dagger Y)_{ii}}{8\pi H(M_i)},
\end{equation}
and the Hubble expansion rate is given by,
\begin{equation}
    H(z)=1.66\sqrt{g_{\star}}\frac {M_1^2}{M_p}\frac {1}{z^2},
\end{equation}
where $g_{\star}=106.75$ and Planck constant $M_p=1.22\times 10^{19}\,\mathrm{GeV}$.

\begin{figure}[b!]
\includegraphics[width=0.5\hsize]{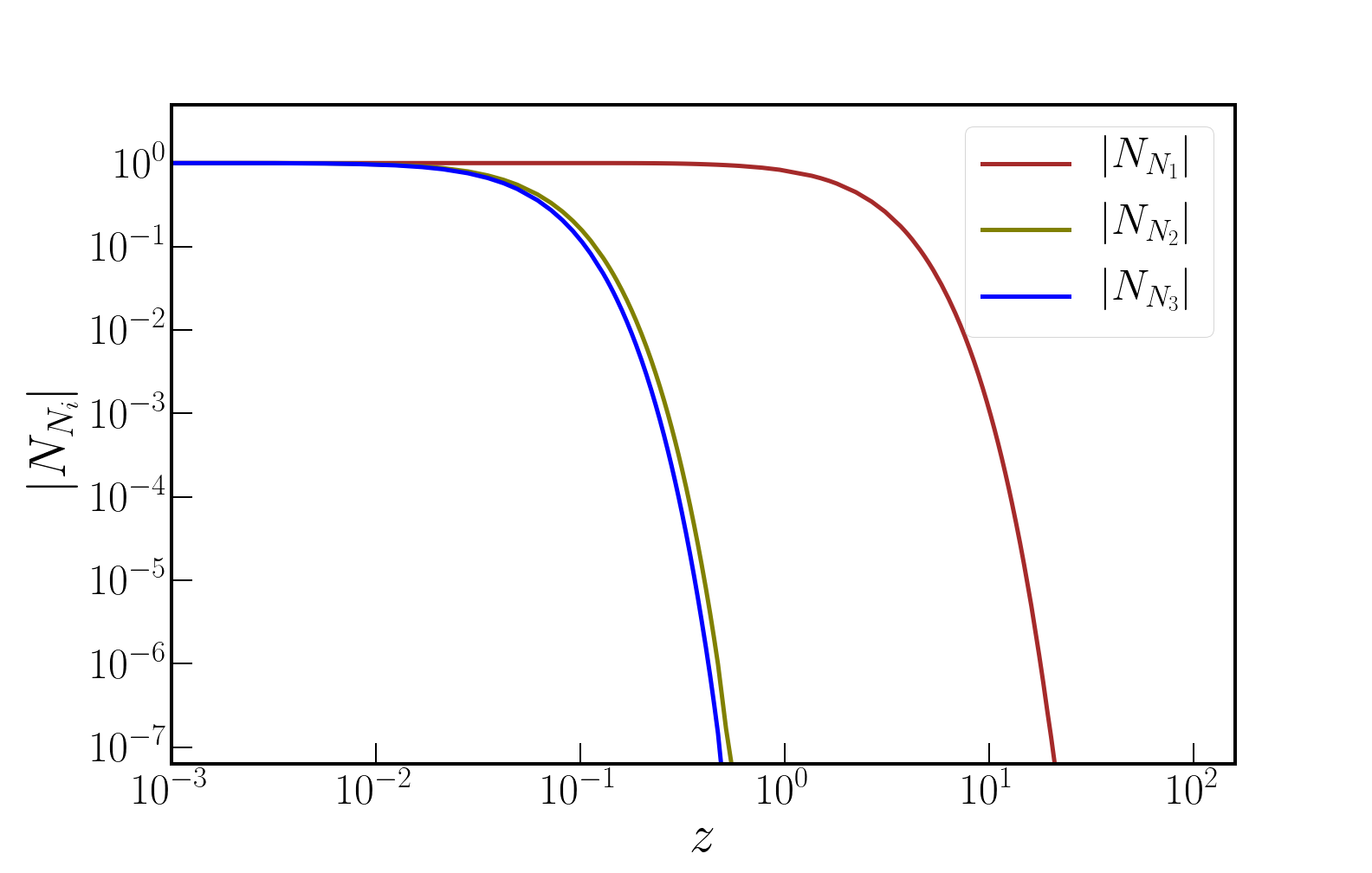}
\includegraphics[width=0.5\hsize]{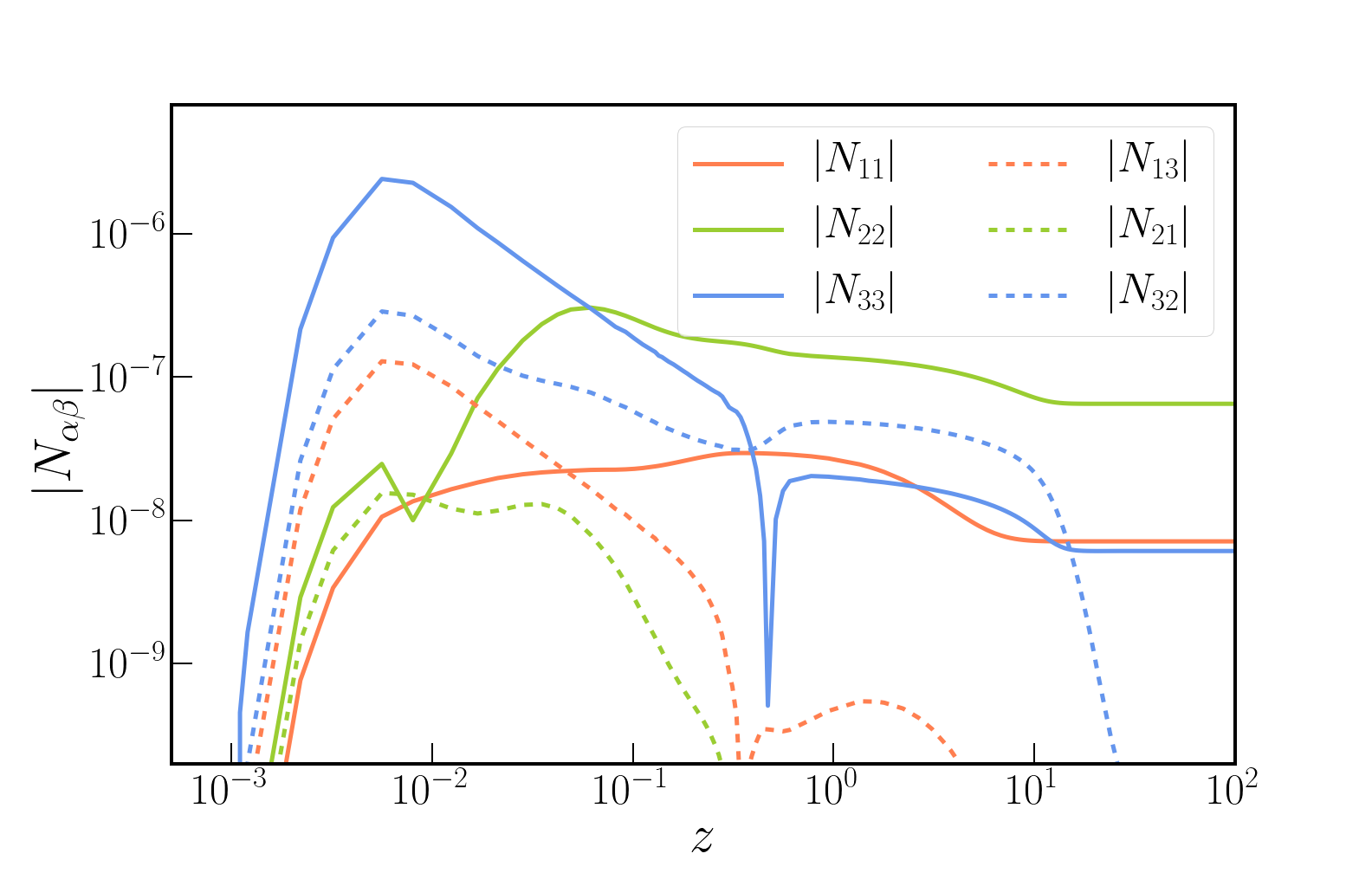}
\caption{The evolution of number densities of right-handed neutrinos $N_{N_i}$ (left panel) and flavored $B-L$ asymmetries $N_{\alpha \beta}$ (right panel) obtained by solving density matrix equations. }
\label{BP1}
\end{figure}

The CP asymmetry matrix $\varepsilon _{\alpha \beta}^{(j)}$ denoting CP asymmetry in the decay of $j^\mathrm{th}$ neutrino in terms of Yukawa coupling and right-handed neutrino masses are given by~\cite{Beneke:2010dz, Blanchet:2011xq},  
\begin{align}
\varepsilon _{\alpha \beta}^{(j)}=\frac {3i}{32\pi (Y^\dagger Y)_{jj}}\sum_{j\neq i}&\left \{ \frac {\xi(x_i/x_j)}{\sqrt{x_i/x_j}}\left [ Y_{\alpha j}Y^*_{\beta j}(Y^\dagger Y)_{ij}-Y^*_{\beta j}Y_{\alpha i}(Y^\dagger Y)_{ji} \right ] \right.  \nonumber\\
&\left. +\frac {2}{3(x_i/x_j-1)}\left [ Y_{\alpha j}Y^*_{\beta i}(Y^\dagger Y)_{ji}-Y^*_{\beta j}Y_{\alpha i}(Y^\dagger Y)_{ij} \right ]\right \},
\label{eq:ephsilon_ab}
\end{align}
with
\begin{equation}
    \xi(x)=\frac 23 x\left [ (1+x)\ln\left ( \frac {1+x}{x} \right )-\frac {2-x}{1-x} \right ].
    \label{eq:xi_x}
\end{equation}
Moreover, $P^{i}_{\alpha \beta}$ is the projection matrix describing how a particular combination of flavor  asymmetry gets washed out via $i^\mathrm{th}$ right-handed neutrino, and its tree-level value is given by~\cite{Blanchet:2011xq},
\begin{equation}
    P^{(i)0}_{\alpha \beta}=C_{i\alpha}^0C_{i\beta}^{*0}=\frac {Y_{\alpha i}Y_{\beta i}^*}{(Y^\dagger Y)_{ii}}.
\end{equation}
 The last two terms in \autoref{eq:dme} describe the effect of charged lepton interaction~\cite{Blanchet:2011xq, Cline:1993bd, Weldon:1982bn}
\begin{align}
    \frac {\mathrm{Im}(\Lambda _\mu)}{Hz}&=\frac {8\times 10^{-3}y_{\mu}^2T}{Hz}=1.7\times 10^{-10} \frac {M_P}{M_1},\\
\frac {\mathrm{Im}(\Lambda _\tau)}{Hz}&=\frac {8\times 10^{-3}y_{\tau}^2T}{Hz}=4.7\times 10^{-8} \frac {M_P}{M_1}.
\end{align}
The $y_\tau$ dependent interaction comes into thermal equilibrium when the temperature drops below $10^{12} \,\mathrm{GeV}$, leading to the decoherence of $\tau$ lepton states. A similar effect arises for $y_\mu$ dependent interaction when the temperature drops below $10^9\, \mathrm{GeV}$. The effect arising from $y_e$ dependent interaction needs to be considered if one considers $M_N<10^6\,\mathrm{GeV}$.

The density matrix is solved numerically, and final $B-L$ asymmetry at $z\gg 1$ can be obtained by taking the trace of $N^{B-L}$ matrix,
\begin{equation}
N^{\mathrm{f}}_{B-L}=\sum_{\alpha}N^{B-L}_{\alpha \alpha}.
\end{equation}
Finally, baryon to photon ratio accounting sphaleron conversion and photon dilution is given by~\cite{Harvey:1990qw, Khlebnikov:1988sr}, 
\begin{equation}
\eta_B=0.96\times 10^{-2}N^{\mathrm{f}}_{B-L}.
\end{equation}
The experimentally measured value of this quantity by Planck~\cite{Planck:2018vyg} and the fit value are summarized in Table~\ref{result}. Moreover, the evolution of the relevant number densities obtained by solving density matrix equations for the benchmark fit presented in Appendix~\ref{A} are depicted in Fig.~\ref{BP1}.

\section{Conclusions}\label{sec:conclusions}
The simplified Yukawa sector within the $SO(10) \times U(1)_\mathrm{PQ}$ framework has garnered considerable attention and has been extensively explored in the existing literature. However, an examination of the renormalization group equations governing the Yukawa couplings that include the threshold effects of the right-handed neutrinos reveals some discrepancy with observed fermion masses and mixings. To address this tension, we proposed an extension (with lower dimensional representations) of the minimal setup by introducing only a few new parameters. The particle content is enlarged to include a fermion in the fundamental representation and a scalar in the spinorial representation. While the latter is crucial for successfully implementing the Peccei-Quinn symmetry, the simultaneous presence of both fermion and scalar proves essential in achieving an excellent fit to the fermion mass spectrum. Furthermore, within our model, the Peccei-Quinn symmetry solves the strong CP problem, and the axion plays the role of dark matter. Additionally, the out-of-equilibrium decays of right-handed neutrinos effectively generate the matter-antimatter symmetry observed in the Universe. This comprehensive approach addresses various challenges, making our proposed model a compelling candidate for reconciling the observed fermion mass spectrum and cosmological phenomena.

\section*{Acknowledgments}
 AK thanks the U.S.~Department of Energy for the financial support under grant number DE-SC 0016013.

\begin{appendices}
\renewcommand\thesection{\arabic{section}}
\renewcommand\thesubsection{\thesection.\arabic{subsection}}
\renewcommand\thefigure{\arabic{figure}}
\renewcommand\thetable{\arabic{table}}

\section{Fit parameters}\label{A}
In this appendix, we provide the fit parameters at the GUT scale for the benchmark solution,
\begin{align}
&r= 17.5677, \;\;\; s= 0.459078+7.79402\times 10^{-3} i, \\
&c_R= 1.90423\times 10^{14},\;\;\; r_0=0.252926, \\
&(r_1,r_2,r_3)=(-2.89426 + 2.13375 i, -1.99196 + 6.79607\times 10^{-4} i, -16.6817 + 1.30852 i),\\
&H=10^{-2} \left(
\begin{array}{ccc}
 3.04492\times 10^{-4} & 0 & 0 \\
 0 & 5.95467\times 10^{-3} & 0 \\
 0 & 0 & 2.59822 \\
\end{array}
\right), \\
&F=
10^{-4}
\left(
\begin{array}{ccc}
 0.378751\, +0.0307759 i & 0.438267\, +0.568287 i & 0.00379929\, -5.50906 i \\
 0.438267\, +0.568287 i & -1.51393+1.17329 i & 5.21488\, -8.68449 i \\
 0.00379929\, -5.50906 i & 5.21488\, -8.68449 i & 1.63784\, -2.7498 i \\
\end{array}
\right).
\end{align}
From the above parameter set, it can be inferred that since $|r_3|\sim 16$, with $z_3\sim 1$ (recall, $\mu_i \lesssim v_{16}$), $m^\prime_F\sim v_{16}/16$ and  $m^{\prime\prime}_F\sim v_{16}/4$. Therefore, heavy vectorlike states have masses, $m_\mathrm{VLF}\sim v_{16}$.

To compute the $\eta_B$ parameter, one needs the right-handed neutrino mass spectrum as well as the Dirac neutrino Yukawa coupling matrix. Masses of the right-handed neutrinos are given in \autoref{eq:RHN}. The formulation in Sec.~\ref{sec:leptogenesis} is performed in the usual $\overline f_L M_f f_R$, which requires $Y_{\nu_D}\to Y_{\nu_D}^\ast\equiv Y$. For the convenience of the readers, here we provide this matrix,   
\begin{align}
Y=\left(
\begin{array}{ccc}
 0.000137864\, +0.000494141 i & -0.00410086-0.00397288 i & 0.00499931\, -0.00497973 i \\
 -0.000337343-0.00322682 i & 0.0082399\, +0.00721626 i & -0.016119+0.00719816 i \\
 0.00788373\, -0.003674 i & -0.04672+0.0919091 i & -0.0952837-0.0648255 i \\
\end{array}
\right),   
\end{align}
given in the charged lepton and right-handed neutrino mass diagonal basis.

Here, we exhibit that the effective $3\times 3$ mass matrix obtained in our derivation is in excellent agreement with the  the full $4\times 4$ matrices. For this demonstration, we consider the down-type quark mass matrix (this can be trivially repeated for the rest of the sectors). Using the fitted values of the parameters, the  $3\times 3$ effective mass matrix is given by (see \autoref{eq:1d}),
\begin{align}
&M_d^\mathrm{light}=
\left(
\begin{array}{ccc}
 0.0162288\, -0.0159864 i & 0.00788095\, -0.042566 i & -0.732447-0.379205 i \\
 0.00661438\, -0.0197767 i & -0.0266143-0.0350817 i & -0.392028+0.17998 i \\
 -0.00197916+0.0124863 i & 0.0114085\, +0.0215266 i & 0.4898\, -0.0268873 i \\
\end{array}
\right),   
\end{align}
whihc has eigenvalues:
\begin{align*}
\left( 1.206 \times 10^{-3}, 2.290 \times 10^{-2}, 1.054 \right)\; \mathrm{GeV}. 
\end{align*}

Next we consider the full $4\times 4$ mass matrix \autoref{full},
\begin{align}
&M_D=
\left(
\begin{array}{cccc}
 0.00712433\, +0.000535821 i & 0.00763041\, +0.0098941 i & 0.0000661471\, -0.095915 i &
   0 \\
 0.00763041\, +0.0098941 i & -0.0159907+0.0204274 i & 0.0907931\, -0.1512 i & 0 \\
 0.0000661471\, -0.095915 i & 0.0907931\, -0.1512 i & 4.55212\, -0.0478751 i & 0 \\
 (-2.89426+2.13375 i) m^\prime_F & (-1.99196+0.000679607 i) m^\prime_F &
   (-16.6817+1.30852 i) m^\prime_F & m^\prime_F \\
\end{array}
\right),   
\end{align}
with eigenvalues:
\begin{align*}
\left( 1.206 \times 10^{-3}, 2.290 \times 10^{-2}, 1.054, 17.259\; \frac{m^\prime_F}{\mathrm{GeV}} \right)\; \mathrm{GeV}. 
\end{align*}
Varying $m^\prime_F$ only changes the mass of the heaviest state, as expected (except for the exception, when $m_F^\prime \lesssim$ TeV, as aforementioned). Since the fit dictates,  $m^\prime_F\approx v_{16}/17$, the vector-like fermion resides at the $v_{16}$ scale.

\section{Expression for $b_0$}\label{B}
The quantity $b_0$ appearing in \autoref{eq:b0} is defined as follows:
\begin{align}
&b_0=b_1+b_2 v^2_7+ v^4_7+ 12 v^2_2 b_3+ v^2_1 b_4,
\\
&b_1=36 v_2^4+36 v_4^4+16 v_5^4+36 v_3^2 v_4^2+4 v_3^2
   v_5^2-48 v_4^2 v_5^2+25 v_3^2 v_6^2,
\\
&b_2=v_3^2-12 v_4^2+8 v_5^2,
\\
&b_3=3 v_3^2+6 v_4^2-4 v_5^2-v_7^2,
\\
&b_4=36 v_2^2+36 v_4^2+4 v_5^2+25 v_6^2+v_7^2.
\end{align}

\end{appendices}

\bibliographystyle{style}
\bibliography{reference}
\end{document}